\documentclass{PoS}
\usepackage{pstricks,pst-grad,color,shadow,graphicx,amsmath,amssymb,amsfonts}

\usepackage{exscale}
\usepackage{slashed}
\usepackage{accents}
\usepackage{bbm}

\let\normalcolor\relax

\newcommand{\psb}{\bar{\psi}}
\newcommand{\ve}{\varepsilon}
\newcommand{\veb}{\bar{\ve}}
\def\pa{\partial}

\def\half{\frac{1}{2}}
\def\Wb{\bar{W}}

\newcommand{\de}{\delta}
\newcommand{\phb}{\bar{\phi}}

\newcommand{\ord}{\mathcal{O}}

\title{Supersymmetric lattice models in one and two dimensions}

\ShortTitle{Supersymmetric lattice models in one and two dimensions}

\author{\speaker{Tobias K\"astner}, Georg Bergner, Sebastian Uhlmann, Andreas
Wipf, Christian Wozar\\
        Theoretisch-Physikalisches Institut,
Friedrich-Schiller-Universit{\"a}t Jena, Max-Wien-Platz 1, 07743
Jena, Germany\\
        E-mail: \email{T.Kaestner@tpi.uni-jena.de}}

\abstract{We study and simulate $\mathcal{N}=2$ supersymmetric Wess-Zumino models in one
and two dimensions. For any choice of the lattice derivative, the theories
can be made manifestly supersymmetric by adding appropriate improvement terms
corresponding to discretizations of surface integrals. In particular, we check that fermionic and bosonic masses coincide
and the unbroken Ward identities are fulfilled to high accuracy. Equally good
results for the effective masses can be obtained in a model with the SLAC
derivative (even without improvement terms). In two dimensions we introduce
a non-standard Wilson term in such a way that the discretization errors of
the kinetic terms are only of order $\ord(a^2)$. Masses extracted from the
corresponding manifestly supersymmetric model prove to approach their continuum
values much quicker than those from a model containing the standard Wilson term.
Again, a comparable enhancement can be achieved in a theory using the SLAC derivative.}

\FullConference{The XXV International Symposium on Lattice Field Theory\\
		 July 30-4 August 2007\\
		 Regensburg, Germany}
\begin{document}

% ===========================================================
% ===========================================================
\section{Introduction}
\noindent Supersymmetry is nowadays an important ingredient in most theoretical
developments of
quantum field theory beyond the standard model. It allows for the unification
of the three fundamental forces described by the standard model and is
also incorporated in supergravity and string theory. In the low energy
regime this symmetry is obviously not manifest and the question remains by which
mechanism supersymmetry if realized in nature is broken. From
non-renormalization theorems it is at least known that this has to be answered
non-perturbatively. In this view the lattice might serve as an equally good
approach as it has been before for gauge theories. However since supersymmetry
is an extension of the Poincar\'e symmetry of spacetime it is inherently broken on
a spacetime lattice.

Here we study and simulate $\mathcal{N}=2$ Wess-Zumino models in one and two
dimensions. Lattice theories with different lattice derivatives
and discretization prescriptions which preserve parts of the supersymmetry are simulated.
It is checked that fermionic and bosonic 
masses coincide and that unbroken Ward identities are fulfilled to high accuracy.
By introducing a nonstandard Wilson term in the two-dimensional theory we can
suppress common $\ord(a)$ artifacts. To include dynamical fermions several
algorithms are used and compared with each other.
For a more thoroughfull presentation of our results we like to refer the reader
to \cite{Bergner:2007pu}.

\section{Quantum Mechanics}
\noindent In the continuum, the action of our first model is given by the action
\begin{equation}S_{\rm cont} = \int d\tau \Big(\frac{1}{2}\dot{\phi}^2 + \frac{1}{2}W'^2
    + \psb\dot{\psi} + \psb W''\psi\Big)\quad\text{with}\quad W'(\phi)\equiv\frac{dW(\phi)}{d\phi};
\end{equation}
it is invariant under the following supersymmetric variations:
\begin{equation}\begin{array}{c@{\qquad}l@{\qquad}l}
\delta^{(1)}\phi = \bar{\ve}\psi,& \delta^{(1)} \psb = -\bar{\ve}(\dot{\phi}+W'), & \delta^{(1)}\psi =0,\\
    \delta^{(2)}\phi = \psb\ve, & \delta^{(2)}\psi =
    (\dot{\phi}-W')\ve, &\delta^{(2)}\psb = 0. 
\end{array} \end{equation}
In order to perform numerical simulations and compare with previously 
results \cite{Catterall:2000rv} we have fix the
potential to \begin{equation}W(\phi)=\frac{m}{2}\phi^2
+ \frac{g}{4}\phi^4.\end{equation}
% ===========================================================

A lattice version of this supersymmetric continuum theory raises a couple of questions.
First we can ask whether the lattice model admits part of the continuum
  supersymmetry. Integrating out the fermions $\psi$ and changing variables from
  the bosons $\phi$ to the so-called Nicolai variables \begin{equation}
\xi=\dot\phi+W'
\end{equation}
renders the bosonic continuum path integral purely Gaussian. Discretizing this sum of squares 
\begin{equation}
\begin{aligned}
S_{\rm bos}=\frac{1}{2}\sum_x \xi_x\, \xi_x&=\frac{1}{2}\sum_x
\left((\partial\phi) + W'(\phi)\right)_x^2 \\
&=S_{\text{naive}}+ \sum_x (\partial\phi)_x W'(\phi)_x,
\end{aligned}
\end{equation}
 one easily verifies that one of the symmetries is preserved.
Since the presence of the additional ``surface'' term improves the behavior of the action
with respect to supersymmetry this action will be called \emph{Nicolai improved}.

Second we have investigated whether there is an optimal lattice prescription for
the Dirac operator. In particular it is a well-known fact that (ultra-)local
hermitean Dirac operators will introduce
fermionic doublers thus spoiling the balance between bosonic and fermionic
degrees of freedom. Two strategies might be pursued, to double the
bosonic spectrum as well or to use the non-local SLAC derivative. The former
requires then to amend the superpotential with a corresponding Wilson term
while the latter is free of any such modifications.

% ===========================================================
\subsection{Degeneracy of mass spectra}
\begin{figure}
\includegraphics[width=0.45\linewidth]{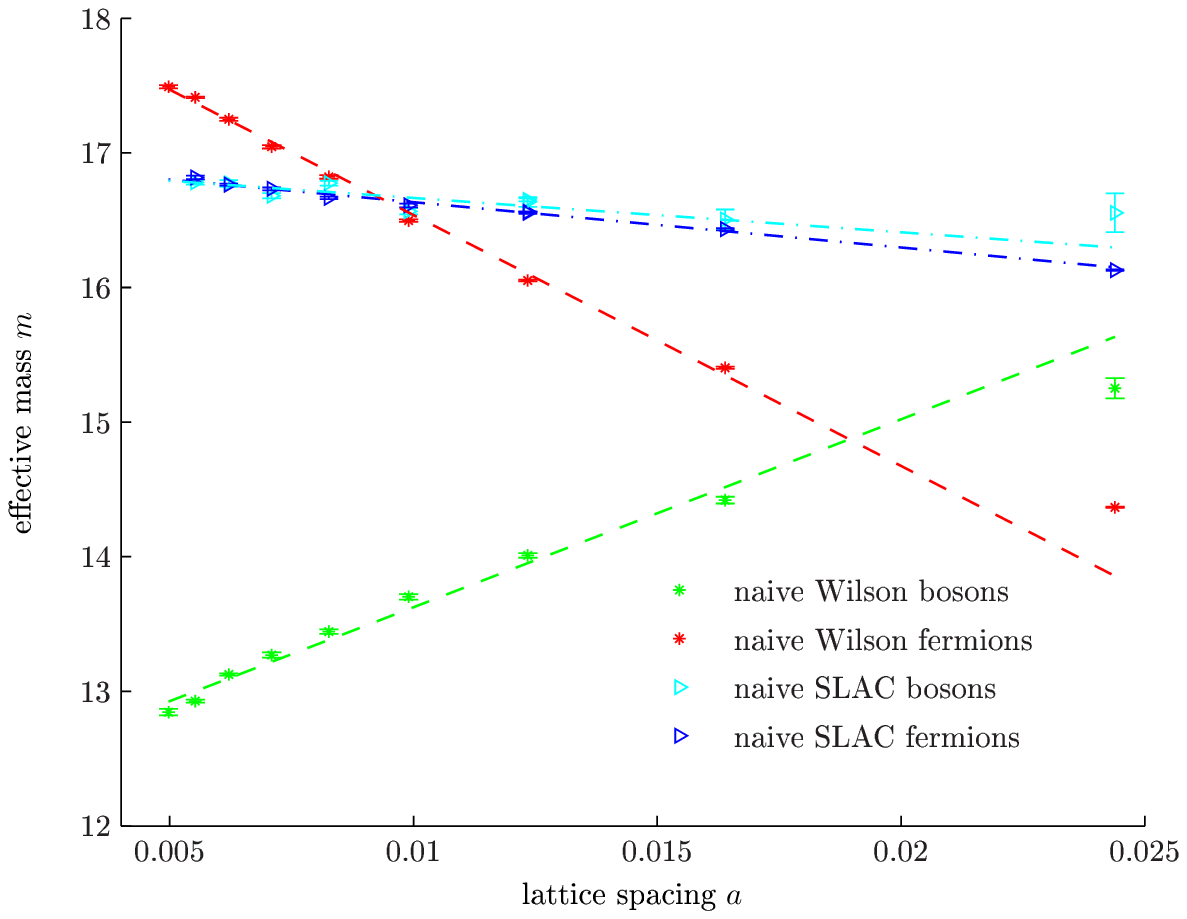}\qquad
\includegraphics[width=0.45\linewidth]{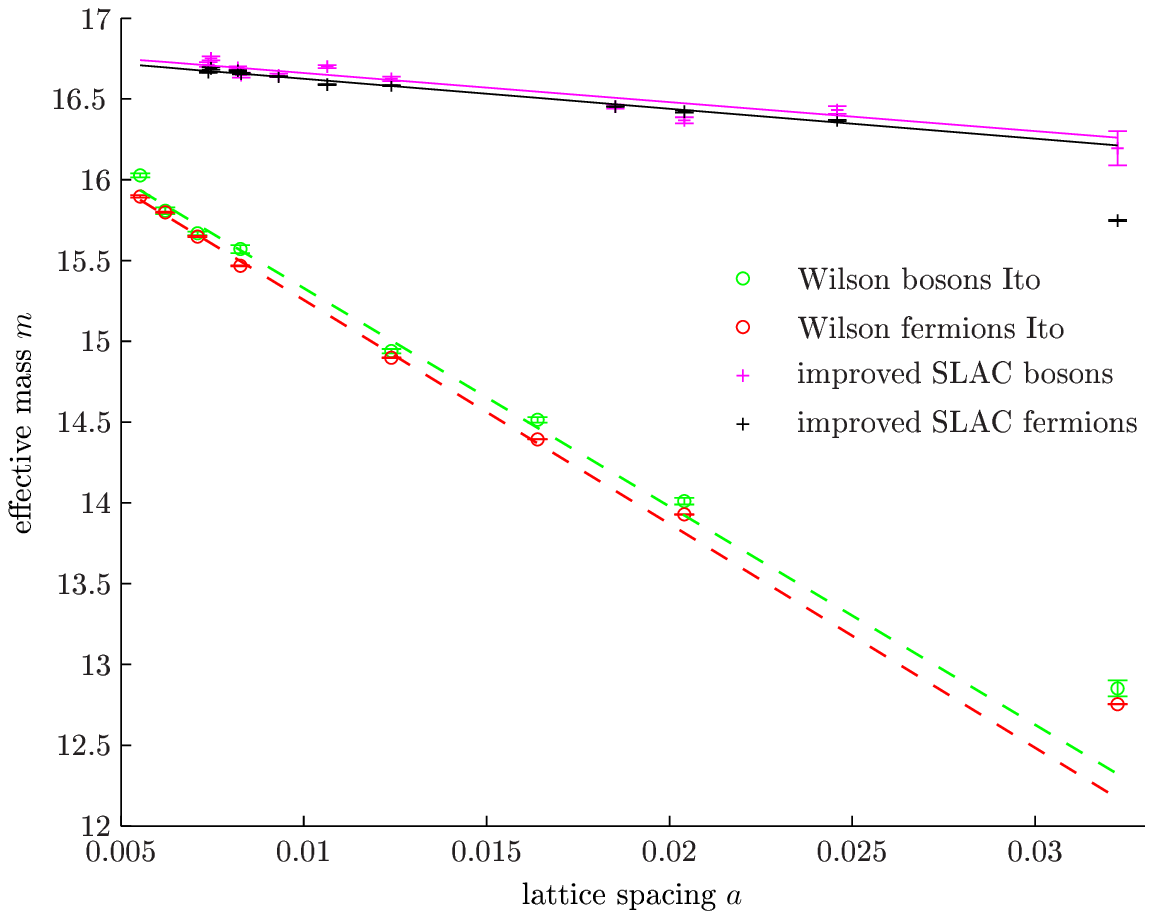}
\caption{ \textbf{Left: }The naive discretization of the continuum action
fails to recover the correct supersymmetric continuum limit when Wilson fermions
are used (green and red graphs). For every finite lattice spacing the extracted
masses differ vastly from each other. The situation is different when SLAC
fermions are used. Here the masses coincide within statistical bounds and tend
towards the correct $a\to 0$ limit. \textbf{Right: }The same graph as on the
left, now using the improved actions as discussed in the main text. Here also
the model with Wilson fermions exhibits degenerate masses at finite lattice spacing and tends towards the correct 
continuum limit.} 
\label{fig:masses}
\end{figure}

\noindent The most obvious physical consequence of supersymmetric theories is
the
degeneracy of masses between the bosonic and fermionic channels which is simply
due to the fact that supersymmetry transforms corresponding states into each
other. In Monte-Carlo simulations the masses of the lowest lying state can be
read off from the
exponential decay of the connected twopoint function which can be readily
measured. For various lattice spacings $a$ we have measured the masses for all
models in both channels, see Fig. \ref{fig:masses}. For all improved actions the presence of
one unbroken
supersymmetry suffices to find the degeneracy even at finite lattice spacing.
However while naive Wilson fermions fail to recover the correct
continuum limit as expected \cite{Giedt:2004vb} and are still plagued by strong $\ord(a)$ artifacts
for the
improved action, SLAC fermions show considerably smaller deviations for finite
$a$ and are much less sensitive to improvement terms.

% ===========================================================
% ===========================================================
\subsection{Ward identities}
\begin{figure}
\includegraphics[width=0.43\linewidth]{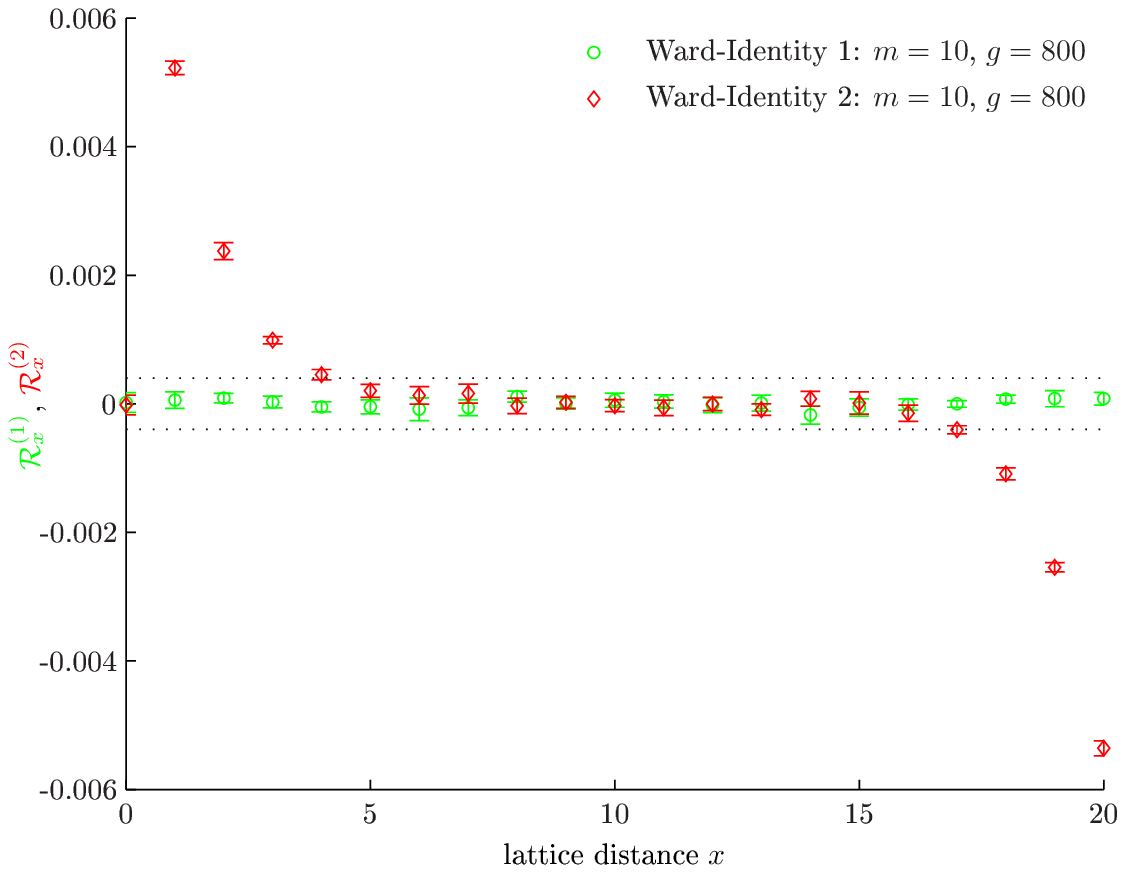}\qquad
\includegraphics[width=0.43\linewidth]{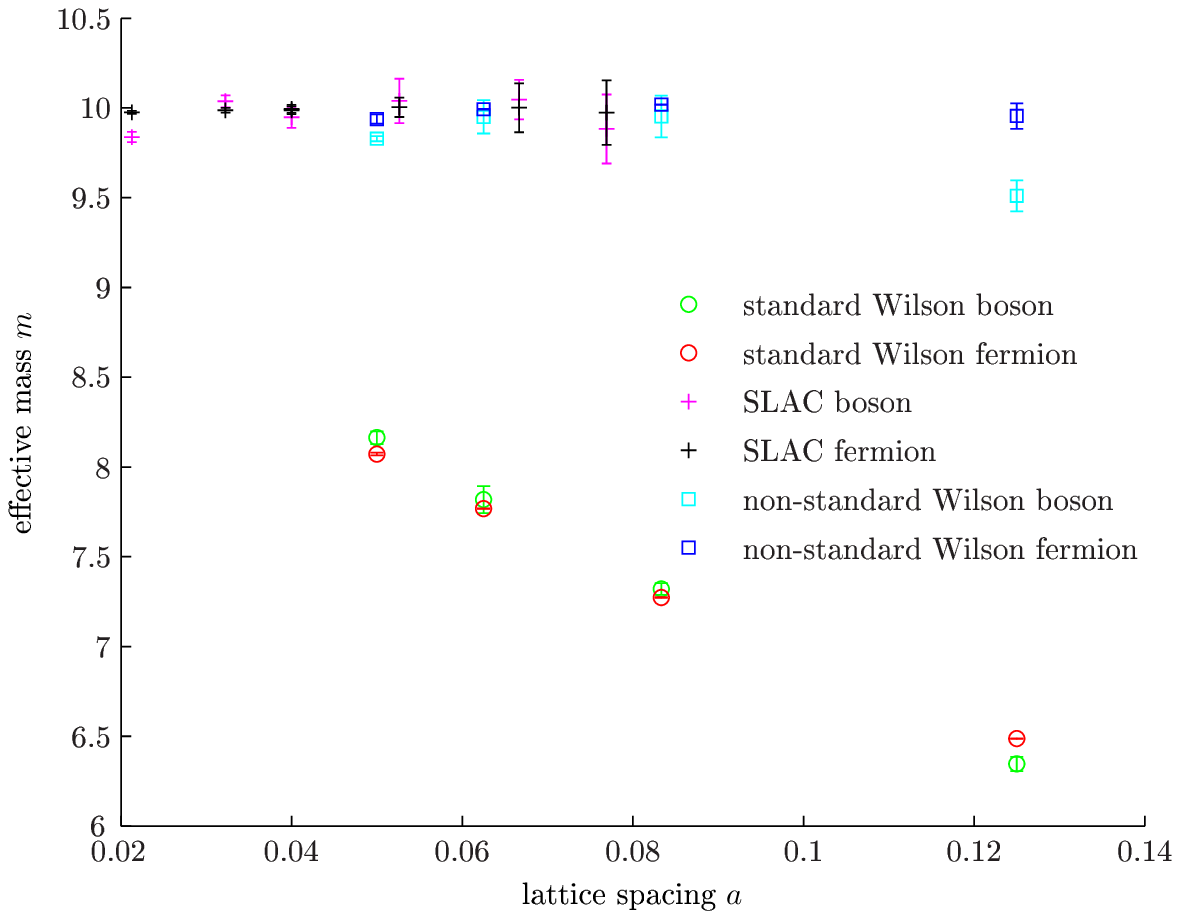}
\caption{
\textbf{Left: }The Ward identities mentioned in the text as a function of $|x-y|$. The
green graph shows the ``intact'' identity while in red the deviation from the
continuum result is clearly visible. The data were taken from a 
lattice with $20$ lattice points using Wilson fermions. The analysis of SLAC
fermions is somewhat more
involved and all signals are worse by an order of magnitude than what is shown
here.  \textbf{Right: }The masses of the various two-dimensional models as a
function of the lattice
spacing. All spectra are degenerate in the scaling region, however lattice
artifacts are much more pronounced for Wilson fermions than for SLAC or modified
Wilson fermions.}
\label{fig:ward}
\end{figure}

\noindent Another important check for the presence of supersymmetry in the
lattice theory is given by the computation of several Ward identities. For any
observable $O$  and supersymmetry variation $\delta$ one should find that
\begin{equation}
\delta\langle O\rangle = \langle \delta O \rangle = 0
\end{equation}
holds. With the particular choice for $O=\phi_x\bar\psi_y$ and $\delta=\delta^{(1)}$
we have checked explicitly the relation
\begin{equation}
\langle\psi_x\bar\psi_y\rangle-\langle\phi_x\xi_y\rangle=
\langle\psi_x\bar\psi_y\rangle-\langle\phi_x(\dot\phi_y+W'_y)\rangle
 =0,
\end{equation}
the results are shown in green on the left of Fig. \ref{fig:ward}.
On the other hand, since $\delta^{(2)}$ is not respected by the lattice action
one might expect 
\begin{equation}
\langle\psi_x\bar\psi_y\rangle-\langle\phi_x(\dot\phi_y-W'_y)\rangle=\langle\delta^{(2)}S\rangle\neq
0 \end{equation}
to hold, cf. the left of Fig. \ref{fig:ward}, too.

By considering Ward identities we have seen that indeed one
supersymmetry is preserved while the other is clearly broken at finite lattice
spacing. Moreover the breaking of Ward Identities 
vanishes rapidly with decreasing lattice spacing and weak
coupling $g$.

 % ===========================================================
\section{Wess-Zumino model in two dimensions}
\noindent The action we start from now reads
 \begin{equation}S_{\rm cont} = 
\int d^2x\Big(
    2 \bar\pa\bar\phi \pa\phi + \half|W'|^2
 +\bar\psi M\psi\Big),\end{equation}
where $W'$ denotes the first derivative of the holomorphic
superpotential and $M$ is given by
\begin{equation}M=\slashed{\pa} +W''P_++\bar W''P_-,\quad P_\pm=\half(\mathbbm{1}\pm
\gamma_3),\quad \psi=\begin{pmatrix}\psi_1 \\ \psi_2\end{pmatrix}. \end{equation}
Again this model is invariant under the following set of supersymmetric variations:
\begin{equation}
\begin{array}{rl@{\quad}rl}
\de\phi & = \psb^1\ve_1 + \veb_1\psi^1, & \de\phb & = \psb^2\ve_2 +
\veb_2\psi^2\\ 
\de\psi^{1} & = -\frac{1}{2}\Wb'\ve_1+\bar{\pa}\phi\ve_2,&\de\psb^1 & =
-\frac{1}{2}\Wb'\veb_1-\pa\phi\veb_2, \\ 
\de\psi^2 & = \pa\phb\ve_1-\frac{1}{2}W'\ve_2, & \de\psb^2 & = -\bar{\pa}\phb\veb_1-\frac{1}{2}W'\veb_2.
\end{array}
\end{equation}
It is still possible to construct a local Nicolai map given by
\begin{equation}
\xi=2(\bar\partial\bar\phi)+W'\,,\;\bar\xi=2(\partial\phi)+\bar W'
\end{equation}
and the \emph{Nicolai improved} bosonic action is thus 
\begin{equation}
S_{\rm bos} = 
\sum_{x}\Big( 
    2 (\bar\pa\bar\phi)_x (\pa\phi)_x + W'_x(\pa\phi)_x
    +  \Wb'_x(\bar\pa\bar{\phi})_x  +  \half|W'_x|^2\Big)
\end{equation}
while the fermionic part reads
\begin{equation}
S_{\rm ferm} =\sum_{x,y}\bar\psi_x M_{xy}\psi_{y},\quad
M=M_0+ W''(\phi_x)\delta_{xy}\,P_+ + \bar W''(\phi_x)\delta_{xy}\,P_-.\; 
\end{equation}
The chosen superpotential differs from the quantum mechanical one and now reads
\begin{equation}W(\phi)=\frac{m}{2}\phi^2
+ \frac{g}{3}\phi^3.\end{equation}
% ===========================================================

This particular lattice actions leaves one of the four continuum supersymmetries
intact. Unlike in the quantum
mechanical case we consider only improved actions
but choose different realizations of the Dirac operator.
Standard Wilson fermions are certainly a natural choice since they are free of
doublers and ultralocal and hence easy and fast to simulate. However, they suffer
from large $\ord(a)$ discretization errors and in our case  necessitate a
modification of the bosonic kinetic operator as well. SLAC fermions are again 
another choice. In order to obtain reasonable results, we have checked that the
theory remains one-loop renormalizable via an explicit perturbative calculation. A
third option emerges from a modification of the standard Wilson term reading 
\begin{equation}
M_0=\gamma^\mu\partial_\mu+\frac{ar}{2}{\color{red}i\gamma_3}\Delta.
\end{equation}
By this twist it can be shown for the free Dirac operator that
all $\ord(a)$ artifacts vanish and the corrections become $\ord(a^2)$. Moreover
on correllators of spatially averaged operators the corrections become even
$\ord(a^4)$ for the free theory. The right panel of Fig. \ref{fig:ward} shows
the masses of the lightest boson and fermion state respectively as a function of
the lattice spacing. Both SLAC and twisted Wilson fermions are much less
disturbed by lattice artifacts than standard Wilson fermions are, although the
improvement of the action ensures the supersymmetric mass degeneracy in all three
cases.

% {
% \includegraphics[width=0.43\linewidth]{ Vergleich_2D_out}
% \flushleft\tiny{The masses of the various models as a function of the lattice
% spacing. All spectra are degenerate in the scaling region, however lattice
% artefacts are much more pronounced for Wilson fermions than for SLAC or modified
% Wilson fermions }
% }
% ===========================================================
% ===========================================================
\section{Algorithms}
\begin{figure}
\includegraphics[width=0.45\linewidth]{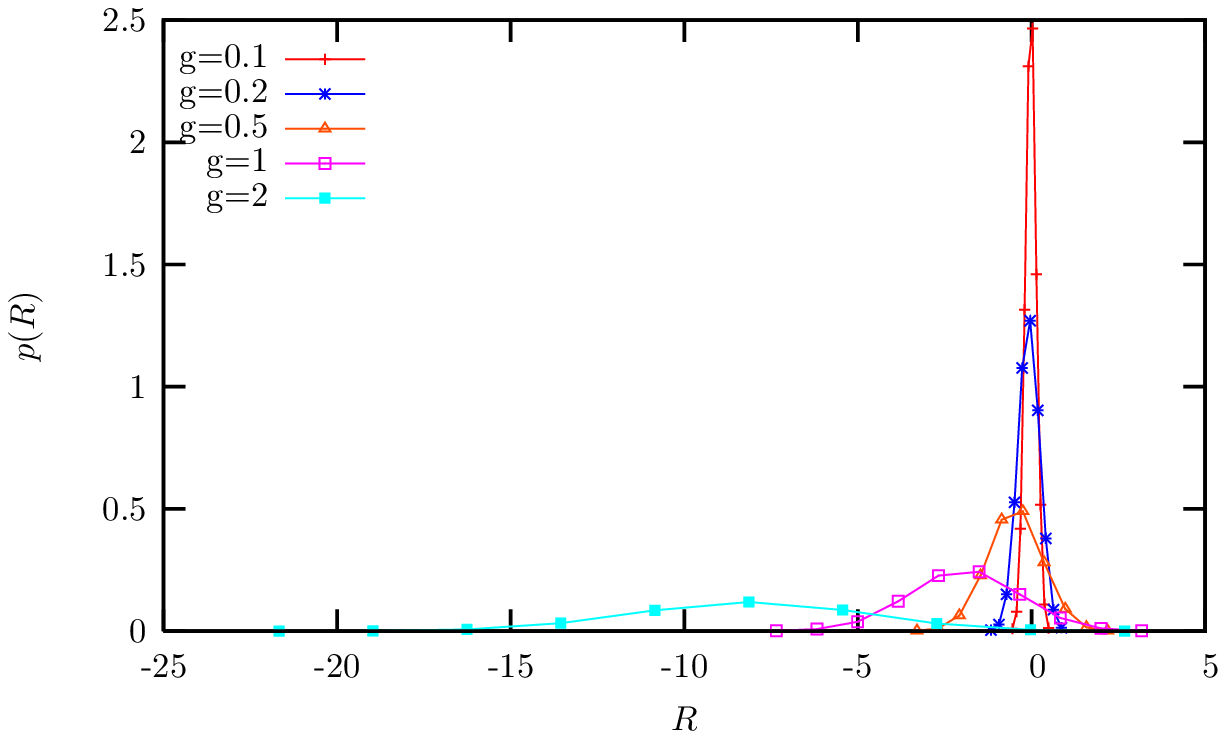}
\quad
\includegraphics[width=0.45\linewidth]{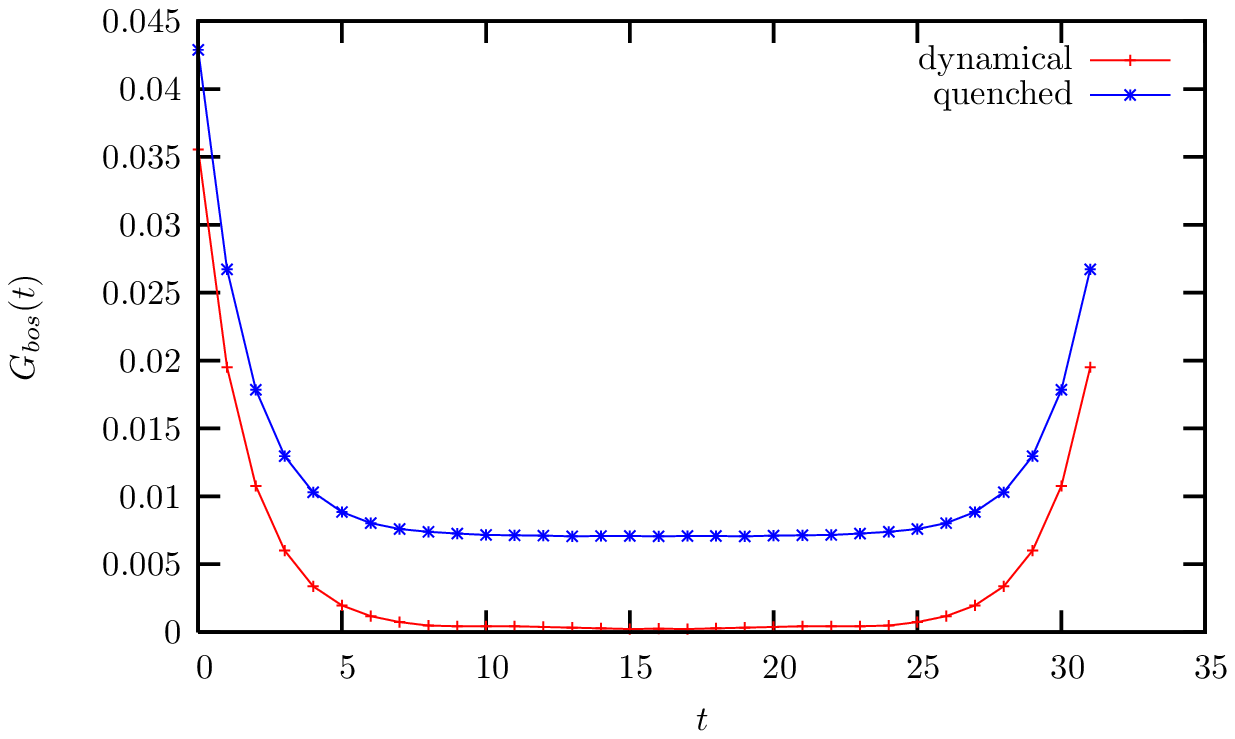}
\caption{
\textbf{Left: }Distribution of the reweighting factor plotted as
logarithm of the fermion determinant normalized to the free field
determinant for different coupling strengths $g$ using SLAC fermions on a
$31\times 31$ lattice. The more pronounced the peak,
the better statistical errors are under control and the more reliably estimates
can be measured. It is obvious that the reweighting technique will fail for
$g\geq 1$. For each distribution 20,000 configurations were evaluated.
\textbf{Right: }Comparison of the bosonic two point function between the
quenched and reweighted ensemble at $g=0.5$ on a $32\times 32$ lattice with
Wilson fermions. The inclusion of fermionic fluctuations in the path integral
are clearly vital for the correct computation of correlation functions and
physical observables.}
\label{fig:reweight}
\end{figure}

\noindent Since low-dimensional theories are less demanding than
four-dimensional LQCD,
several strategies to handle the fermion determinant on top of the standard
hybrid Monte Carlo might be put to use. In any case the models deviate at
least in two points from the more familiar scenario of LQCD. First our theories
involve a only single flavor in order to keep the fermionic an
bosonic degrees of freedom balanced and second, $\gamma_5$-hermiticity is broken by the Yukawa coupling terms.  % \begin{enumerate}
%   \item The theory must use only a \emph{single flavour} in order to keep the number
%   of degrees of freedom balanced.
%   \item The Yukawa like coupling breaks $\gamma_5$-hermiticity, i.e. $\gamma_5
%   D \gamma_5 \neq D^\dagger$. In addition the fermion determinant is in two
%   dimensions \emph{not strictly positive}. 
% \end{enumerate}
%
In view of this and to start the investigation on safe grounds, various
treatments of the fermion determinant are used in parallel and compared to each
other. In the simplest case, the quantum-mechanical model with Wilson fermions,	
the explicit formula for the fermion determinant reads
 \begin{equation}
 \det M_W[\phi] = \prod_x \left( 1+m+3g\phi_x^2\right) - 1,
 \end{equation}
and can be applied directly to include fermionic contributions in a HMC
integration scheme. Since the computational effort is rather small very high
statistics are attainable. In the second related method one computes the
determinant and the inverse of the fermion matrix by direct methods such as 
LU-factorization. Again additional noise originating from the use of 
pseudo-fermions is absent. While easily applicable in one dimension the method
soon becomes infeasible in two dimensions as the lattices grow in size. A third
possibility is given by reweighting the fermionic contribution from quenched
ensembles. This method can generate configurations very quickly and is still
exact in its treatment of fermionic fluctuations. Nonetheless it fails rapidly with
increasing coupling constants since the fluctuations might then overstretch more
than twenty orders of magnitude, see the left of Fig. \ref{fig:reweight}, thereby reducing the
effective number of configurations to order one. Finally pseudo-fermions are a
well-known approach to estimate the fermion determinant stochastically. Recent
algorithms such as PHMC and RHMC allow for the treatment of
fractional powers of $M^\dagger M$. Thus these algorithms can also be used to
simulate supersymmetric single-flavor field theories. However, the annoying
problems with small eigenvalues of the fermion matrix will remain and may hamper
the numerical treatment of these models.% % %\vspace{.5cm} %
%\textbf{\color{blue}1. Closed form expression} % \begin{itemize}
%   \item an explicit expression for the fermion determinant in one dimension
%   using Wilson fermions is known
%  \begin{equation}
%  \det M_W[\phi] = \prod_x \left( 1+m+3g\phi_x^2\right) - 1
%  \end{equation}
%    {\item very fast evaluation}
%    {\item only applicable to Wilson fermions in one dimension}
% 
% \end{itemize}
% 
% \vspace{.5cm}
% \textbf{\color{blue}2. Brute force}
% \begin{itemize}
%   \item computation of the determinant and the inverse of $M$ through LU-factorization
% {\item no additional noise from pseudo-fermions}
% {\item in two dimensions the inversion might become singular due
% to very small eigenvalues }
% \end{itemize}
% 
% \vspace{.5cm}
% \textbf{\color{blue}3. Reweighting}
% \begin{itemize}
%   \item Observables are estimated by reweighting quenched ensembles
% {\item computation of determinant is needed only once per trajectory}
% {\item only weakly coupled regime accessible}
% \end{itemize}

% \vspace{.5cm}
% \textbf{\color{blue}4. Pseudo-fermions}
% \begin{itemize}
%   \item Standard approach for the inclusion of dynamical fermions
% {\item comparable fast iterative solvers or approximations
% (PHMC, RHMC) can be used} 
% {\item not limited to weak coupling
% regime} 
% {\item introduces additional noise}
% {\item again sensitive to instabilities from small and/or
% negative eigenvalues}
% \end{itemize}
% \vspace{0.8cm}

% ===========================================================
\section{Conclusions and Outlook}
\noindent We have tested several lattice constructions of supersymmetric
$\mathcal{N}=2$ 
Wess-Zumino models in one and two dimensions. The extended supersymmetry algebra
admits the construction of a lattice action which preserves one supersymmetry.
Using Wilson fermions this single remnant of the continuum symmetry suffices to
observe important features of the theory such as a degenerate mass spectrum and
the validity of associated Ward identities independent of the choosen lattice derivative.
For the SLAC derivative in one dimension we have found that the results do not
differ vastly between the naive and improved action respectively.
With the help of the derived
Ward identities it is possible to check explicitly that one supersymmetry is
respected while the other is broken.

The two-dimensional models are numerically more demanding since upon integrating
out the fermion fields one ends up with an (in general not strictly positive)
determinant. This situation worsens when the coupling is made stronger leaving
this regime inaccessible for reweighting techniques. However the correct
treatment of fermionic fluctuations is again crucial for the expected
``supersymmetric'' physics to show up as can be seen from the right of Fig. \ref{fig:reweight}.
With the introduction of a modified
Wilson term the typical $\ord(a)$ scaling is circumvented yielding 
results of about the same quality as the non-local SLAC fermions.
% ===========================================================

In order to investigate the whole parameter space and/or models in more
  than two dimensions some technical obstacles related to the treatment of the
  fermion determinant must be readdressed. In particular we know from first
experiments that preconditioning the linear systems before applying iterative solver schemes would lead to a
  significant gain. Furthermore other acceleration techniques such as Fourier
  accelaration, multiple time-scales or higher order integrators are under
  investigation. With the help
  of the PHMC algorithm we hope to extend the stability of the algorithm into
  regions of parameter space which are inaccessible at the moment.
Apart from this, it is already possible to study further supersymmetric
models such as the $\mathcal{N}=1$ Wess-Zumino model in two dimensions, nonlinear supersymmetric $\sigma$-models,
Wess-Zumino models in higher spacetime dimensions and Super-Yang-Mills theories
with the help of our existing codes.

\acknowledgments

\noindent TK acknowledges support by the Konrad-Ade\-nauer-Stiftung
e.V., GB by the ev. Studienwerk Villigst e.V. and CW by
the Studienstiftung des deutschen Volkes.
This work has been supported by the DFG grant Wi 777/8-2.

% ===========================================================

\end{document}